\pdfoutput=1 

\documentclass{JINST}

\title{Characterization of photo-multiplier tubes for the Cryogenic Avalanche Detector }

\author{A.Bondar$^{ab}$, A.Buzulutskov$^{ab}$, A.Dolgov$^{ab}$, V.Nosov$^{ab}$, 
L.Shekhtman$^{ab}$\thanks{Corresponding author}, A.Sokolov$^{ab}$ \\
\llap{$^a$}Budker Institute of
Nuclear Physics SB RAS\\ 11 Lavrentiev Avenue, Novosibirsk 630090\\
Russia. \\ e-mail: L.I.Shekhtman@inp.nsk.su\\
\llap{$^b$}Novosibirsk State University\\
2 Pirogova str., Novosibirsk 630090, Russia\\}

\abstract {New Cryogenic Avalanche Detector (CRAD) with ultimate sensitivity, that will be able to detect one primary electron released in the cryogenic
liquid, is under development in the Laboratory of Cosmology and Particle Physics of the
Novosibirsk State University jointly with the Budker Institute of Nuclear Physics. 
The CRAD will use two sets of cryogenic PMTs in order to get trigger signal either from primary scintillations in liquid Ar or
from secondary scintillations in high field gap above the liquid. Two types of cryogenic PMTs produced by Hamamatsu Photonics were tested and the results are
presented in this paper. 
Low background 3 inch PMT R11065-10 demonstrated excellent performance according to its specifications provided by the producer. The
gain measured with single electron response (SER) in liquid Ar reached 10$^7$, dark count rate rate did not exceed 300 Hz and pulse height resolution of single electron
signals was close to 50\%(FWHM). However, two R11065-10 PMTs out of 7 tested stopped functioning after several tens minutes of operation immersed completely into
liquid Ar.
The remaining 5 devices and one R11065-MOD were operated successfully for several hours each with all the parameters according to the producer specifications. 
Compact 2 inch PMT R6041-506-MOD with metal-channel dynode structure is a candidate for side wall PMT system that will look at
electroluminescence in
high field region above liquid. Four of these PMTs were tested in liquid Ar and demonstrated gain up to 2x10$^7$, dark count rate rate below 100 Hz and pulse height resolution of
single electron signals of about 110\% (FWHM).} 

\keywords
{dark matter, Cryogenic Avalanche Detector, cryogenic PMT}

\begin{document}

\section{Introduction.}
Two-phase Cryogenic Avalanche Detector is proposed and being developed in the Laboratory of Cosmology and Particle physics of the
Novosibirsk State University jointly with the Budker Institute of Nuclear Physics ~\cite{CRAD1, CRAD2, CRAD3}. 
Main purpose of this detector is to search for weakly
interacting massive particles (WIMP)~\cite{DarkM1} and coherent neutrino-nucleus scattering ~\cite{cohneut1, cohneut2} 
by detecting signals from nuclear recoils ~\cite{Recoils}. The energy of recoil nuclei from WIMP is predicted not to exceed several tens keV with the most interesting range
below 8 keV that corresponds to the WIMP mass below 10 GeV. For coherent neutrino-nucleus scattering the recoil energy is even lower, below 0.5 keV. Thus the
CRAD sensitivity that can allow detection of single primary electrons becomes a factor of crucial importance.

\begin{figure}[htb]
\centering
\includegraphics[width=0.7\textwidth,viewport=1 50 500 700,clip]{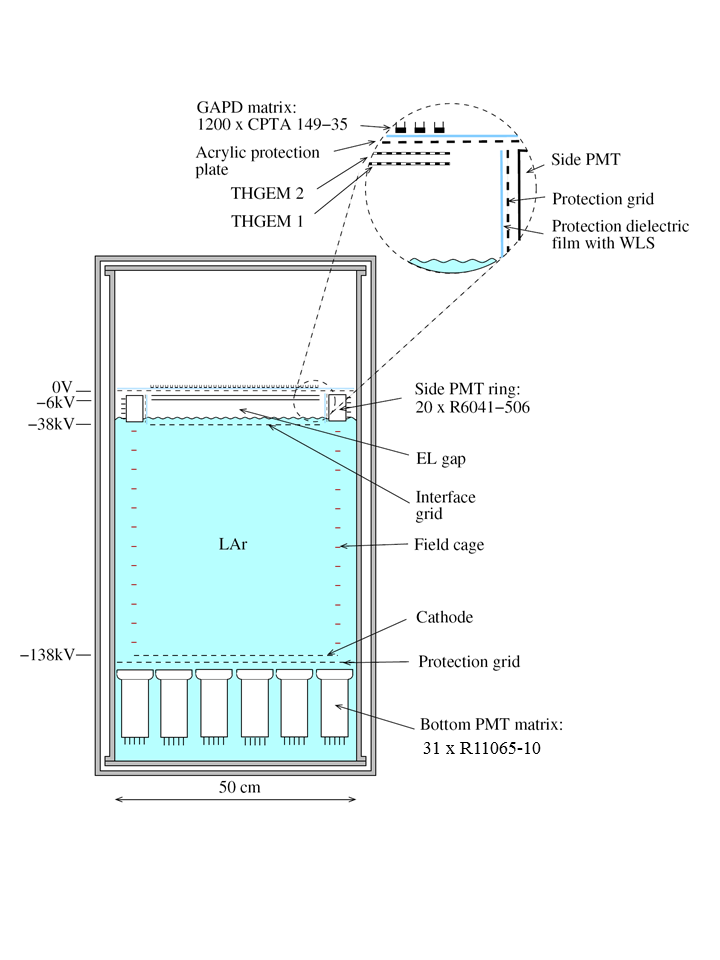}
\caption{Design of two-phase CRAD in Ar with optical readout by combined GEM/GAPD multiplier.}
\label{fig:CRAD}
\end{figure}

The design of proposed CRAD is shown schematically in Fig.~\ref{fig:CRAD}.   
The detector consists of cryogenic vessel with vacuum thermal insulation that can hold up to 150 l of liquid Ar (200 kg). Sensitive volume is limited by
the height of bottom PMT system as well as the thickness of side PMTs and is close to 50 l (50 cm height and 36 cm diameter). Primary electrons, produced by
recoil nuclei after WIMP or neutrino scattering, drift in the liquid towards the surface in electric field of 2 kV/cm. At the liquid surface electrons are
emitted in the gas phase where they produce signal in two stages. Firstly by electroluminescence in high electric field above the liquid, secondly by
avalanche multiplication in double-THGEM cascade or in a hybrid 2THGEM/GEM cascade. Proportional scintillations in VUV region 
from electroluminescence (EL) gap are detected by the ring of side PMT and by the bottom PMT system. To provide sensitivity in VUV range all PMT have
wavelength shifter (tetraphenyl-butadiene, TPB) deposited either directly on the PMT window or on thin transparent plastic foil in front of them. 
Bottom PMT are also intended for detection of primary scintillations in the liquid. 

Bottom and side PMT develop trigger signal for the matrix of Geiger-mode avalanche photo-diodes (GAPD) that detects avalanche scintillation signals from the top
THGEM in the Near Infrared (NIR) ~\cite{CRAD1, CRAD2}. GAPD matrix consists of 3x3 mm$^2$ sensors placed 
in square or hexagonal lattice on a rigid printed circuit board (PCB) at a distance of 1 cm from the top
THGEM. Such arrangement can provide spatial resolution below $\sim$2 mm for single primary electrons ~\cite{THGEM-GAPD1} that is better than in classic 
cryogenic detectors with
only PMT readout. Good spatial resolution together with the possibility to distinguish events with one and two primary electrons allows for effective
rejection of single electron background. This is a unique feature of the proposed design that can be realized only if the PMT trigger system can separate
effectively single primary electron events. Taking into account the number of secondary photons emitted in the EL gap of 4 cm with electric field of 8 kV/cm, 
the solid angle of all PMTs, PMT quantum efficiency and re-emission efficiency of the TPB
layer we can estimate the number of photo-electrons in PMT, corresponding to a single primary electron in the liquid, as 23 ~\cite{CRAD2, CRAD3}. These
photoelectrons will be distributed over more than 50 PMT. Thus it is important that the PMT will detect single electrons with good efficiency (higher than 90\%). 

For the bottom PMT system R11065 Hamamatsu 3 inch PMT was chosen based on the experience of WArP collaboration ~\cite{PMT-WARP}. 
This PMT has bialcali photocathode adapted for operation down to liquid Ar temperature (87 K). Quantum efficiency (QE) 
reported by the manufacturer is not less than 25\% at 420 nm. The
authors of  ~\cite{PMT-WARP} report however, the value of 33\% QE at 420 nm. During 2012 and 2013 Hamamatsu Photonics undertook significant steps to reduce
radioactivity of this PMT and developed R11065-10 and R11065-20 tubes. Replacement of a glass stem with a ceramic one lead to  
significant reduction in $^{40}$K content for both modifications. In R11065-20 cobalt free alloy was used that decreased the amount of $^{60}$Co. In R11065-10 the radioactivity per tube is
17 mBq compared to 75 mBq for R11065 ~\cite{PMTLAr1}. However during 2013 and 2014 it was reported that new modifications of R11065 , in particular
R11065-20, have problems operating in liquid Ar (see for example ~\cite{PMTLAr2}). Thus testing a set of PMT for the bottom system purchased from Hamamatsu, 
7 R11065-10 and 1 R11065-MOD, became an urgent issue. 

For the side PMT system we chose Hamamatsu R6041-506MOD device. This is a compact 2 inch PMT with metal-channel dynode structure and bialcali photocathode adapted for
operation in liquid Ar (87 K) with 25\% QE at 420 nm ~\cite{R6041specs}. This tube has 57 mm diameter and 43.5 mm height, that allows to keep 
the diameter of sensitive volume not less than 36 cm. The report on its successful operation in liquid Ar can be found in ~\cite{R6041appl}, however, with no information on its
operation parameters. Thus we needed to test these PMT to become sure that they can detect  single photoelectrons with probability at least higher than 90\% in liquid Ar.

\section{Experimental set-up and procedures. } 

\begin{figure}[htb]
\centering
\includegraphics[width=0.8\textwidth,viewport=20 50 500 500,clip]{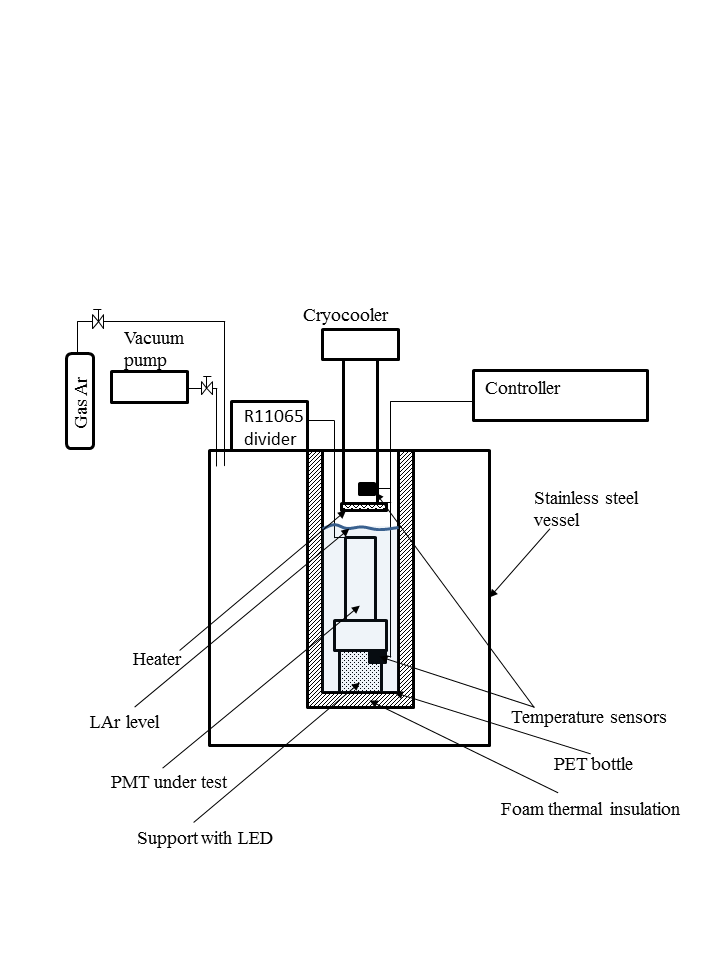}
\caption{Schematic set-up for the PMT tests.}
\label{fig:Design}
\end{figure}

Schematic design of the PMT test set-up is shown in Fig.~\ref{fig:Design}.
For the tests of PMT we used stainless steel vessel 30 cm in diameter and 40 cm high with gas-tight top cover. The vessel can keep up to 3 atm of
overpressure. The top cover contains 3 metal-glass feedthroughs that were used to provide high voltage to the PMT under test, signal from the pulse generator 
to the light emission diode (LED), signals from the temperature sensors to the Lake Shore 336 temperature controller 
and voltage to the heater from the same controller. 

\begin{figure}[htb]
\centering
\includegraphics[width=0.7\textwidth,viewport=1 1 700 500,clip]{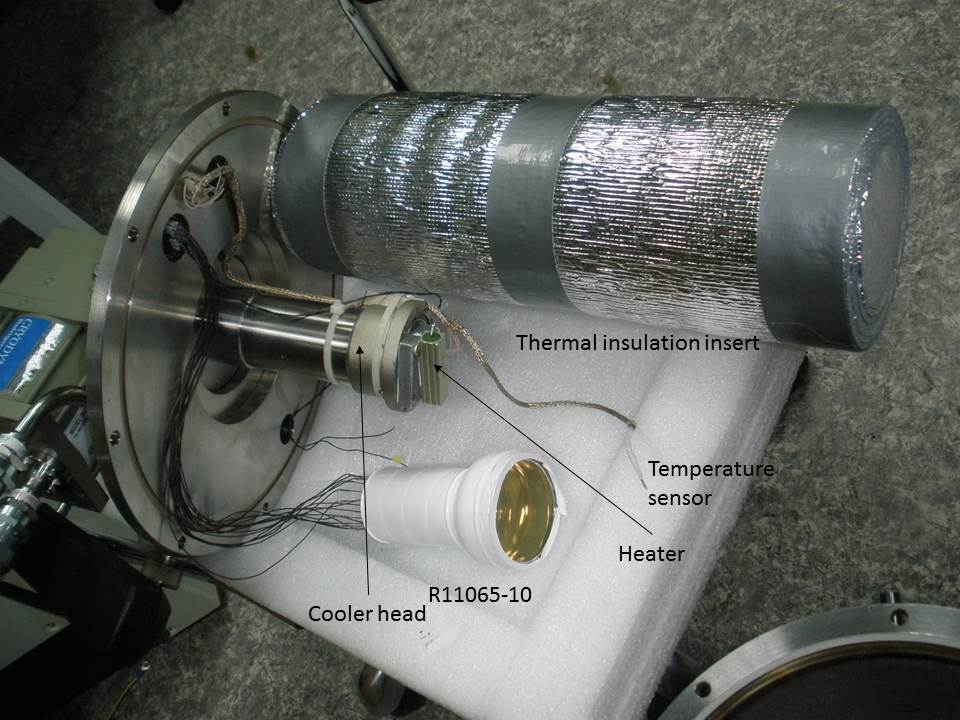}
\caption{Inner assembly of the PMT test set-up.}
\label{fig:Set-upphoto}
\end{figure}

Cryocooler Cryodyne 350CP by CTI-Cryogenics was used for cooling and liquefaction of Ar. 
PMT under test was installed inside a thermal insulating cylindrical case 10 cm in
diameter and 35 cm high. The case has plastic PET (polyethylene terephthalate) bottle $\sim$8cm in diameter and $\sim$30 cm high firmly attached inside. 
At the bottom the bottle has special support containing the LED for PMT
cathode illumination. The whole inner assembly is shown at the photo in Fig.~\ref{fig:Set-upphoto}. 

\begin{figure}[htb]
\centering
\includegraphics[width=0.7\textwidth,viewport=1 1 700 500,clip]{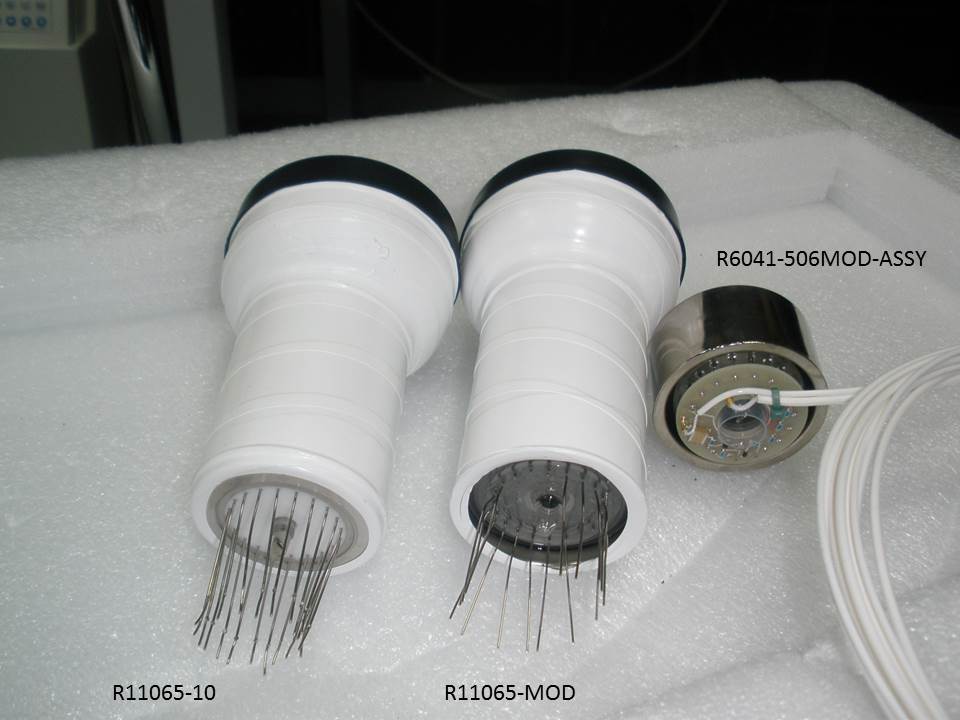}
\caption{Photograph of R11065-10, R11065-MOD and R6041-506-ASSY PMT (from left to right). .}
\label{fig:PMT-photo}
\end{figure}

At the first stage we plan to put in the bottom PMT system seven PMT. Therefore seven R11065-10 and one R11065-MOD were purchased from Hamamatsu. The photo of
both tubes is shown in Fig.~\ref{fig:PMT-photo} where different stem material can be distinguished: ceramic in R11065-10 and glass in R11065-MOD. 15 contacts
from PMT were connected through the metal-glass feedthrough in the top cover to the resistive divider (12 dynodes, ground, anode and grid). The
circuit of the divider can be found in ~\cite{R11065divider}).   

Four R6041-506MOD-ASSY were tested out of 21 purchased from Hamamatsu. This PMT has resistive divider on it (see Fig.~\ref{fig:PMT-photo}). The circuit of
this divider can be found in ~\cite{R6041specs}.

\begin{figure}[htb]
\centering
\includegraphics[width=0.9\textwidth,viewport=1 200 600 600,clip]{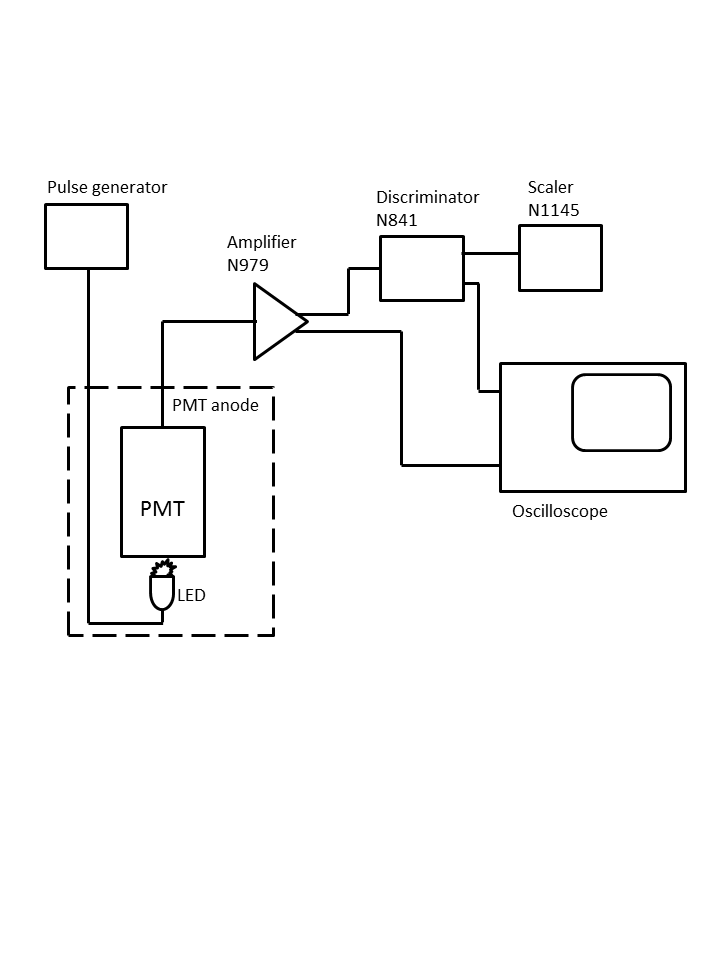}
\caption{Block diagram of the electronics for PMT tests.}
\label{fig:Set-up}
\end{figure}

Both R11065 and R6041-506MOD-ASSY PMTs were powered by positive voltage from N1470 CAEN high voltage power supply. 
Block-diagram of the electronics for PMT tests is shown
in Fig.~\ref{fig:Set-up}.
Signal from PMT anode was amplified by a factor of 10 or 100 by CAEN N979 fast linear amplifier module. Then amplified signal was connected to oscilloscope
Tektronix TDS5034B and to discriminator CAEN N841 for counting rate measurements by scaler CAEN N1145. Oscilloscope was used for the measurements of waveforms and pulse
height spectra. Signals were obtained from spontaneous emission of single electrons from photocathode or by illumination with LED. The LED was powered from
pulse generator G5-78, and its light intensity could be tuned by changing pulse height and base line from the generator. The rate with LED did not exceed 5 kHz and
the frequency of pulse generator was 50 kHz. 

\section{Results and discussion.}
\subsection{Results with R11065-10 and R11065-MOD PMTs.}

\begin{figure}[htb]
\centering
\includegraphics[width=0.7\textwidth,viewport=50 20 800 550,clip]{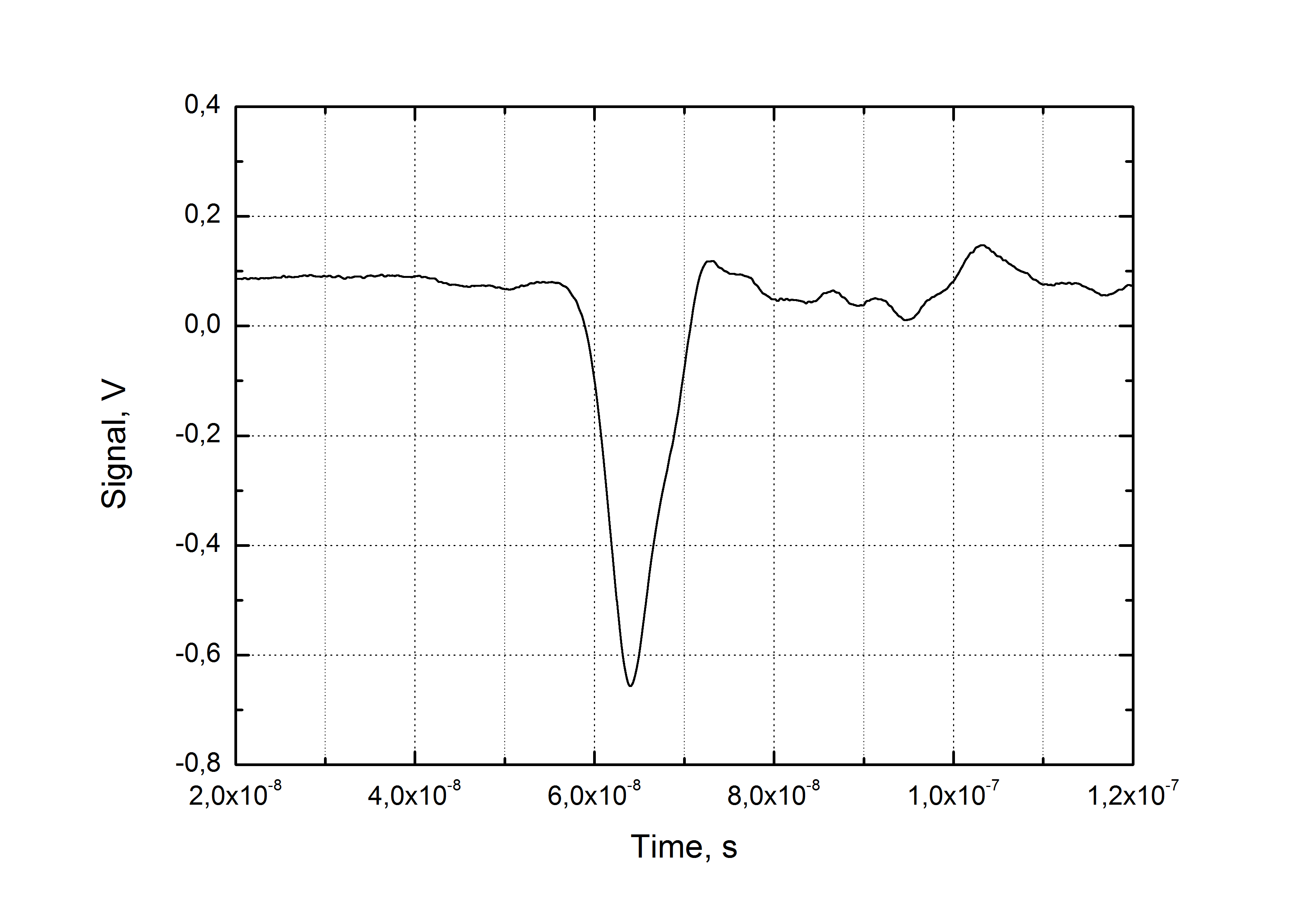}
\caption{Single electron response from R11065-10. Anode voltage: 1500 V, gain: $\sim$10$^7$.The signal is amplified by a factor of 100.  }
\label{fig:R11065shape}
\end{figure}

\begin{figure}[htb]
\centering
\includegraphics[width=0.7\textwidth,viewport=50 400 500 730,clip]{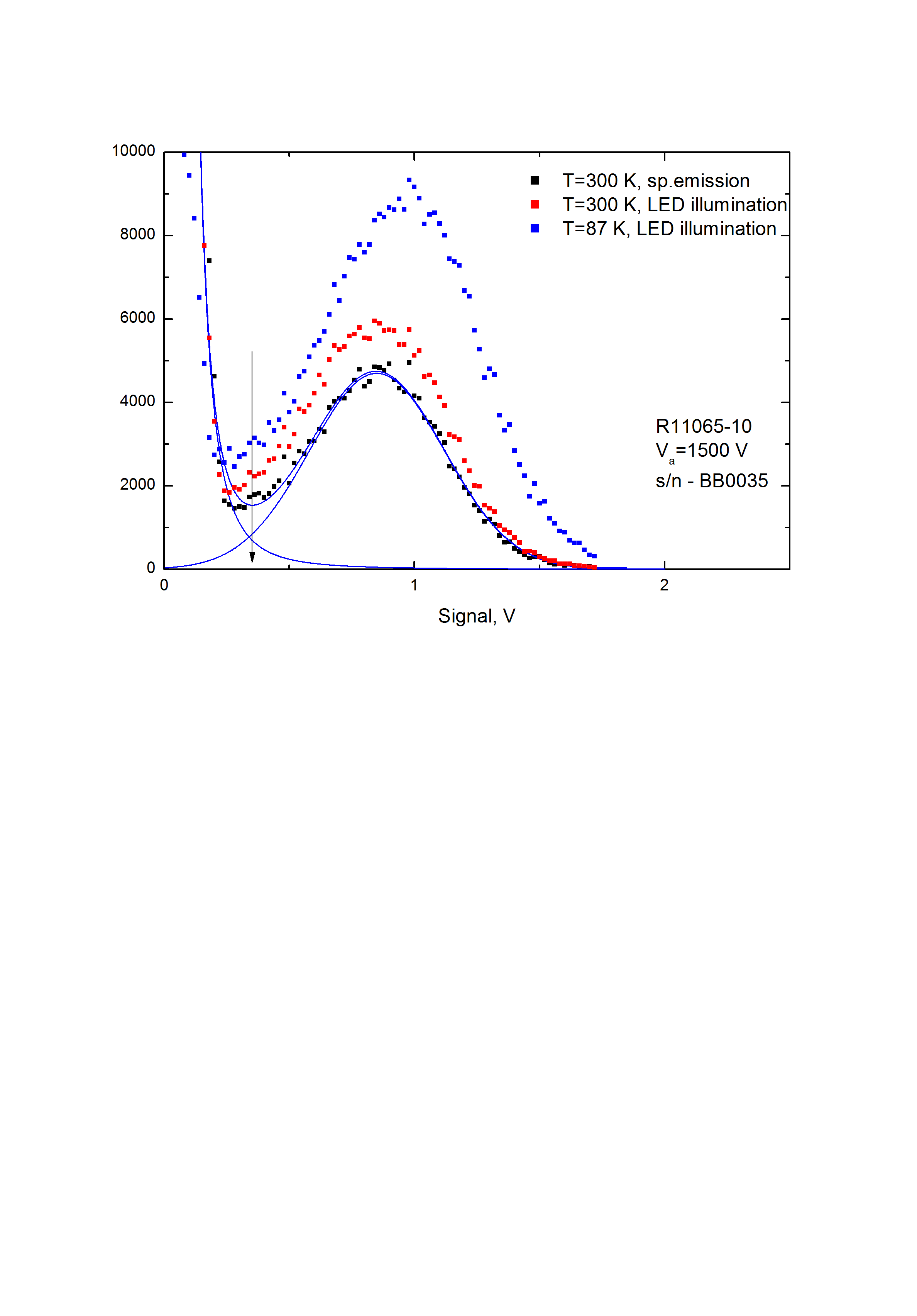}
\caption{Pulse height spectra of SER obtained from spontaneous emission (black points) and with LED illumination at room temperature (red points) and at 87 K (blue points). Vertical
arrow points to the threshold for calculation of the relative single electron efficiency.  
Anode voltage 1500 V, gain $\sim$10$^7$.}
\label{fig:R11065spectrum}
\end{figure}

Pulse shape of single electrons of spontaneous emission in R11065-10 is shown in Fig.~\ref{fig:R11065shape}. The signal is amplified by a factor of 100. The
signal is fast with FWHM of $\sim$7 ns and full width at 0.1 maximum around 12 ns. The pulse height spectrum of SER of spontaneous emission at room
temperature is shown in Fig.~\ref{fig:R11065spectrum} (black points). 
The spectrum is very well fitted by Gaussian distribution without any trace of double-electron peak.

The rate of spontaneous emission is rather low at liquid Ar temperature, in particular for R6041-506MOD-ASSY tubes; therefore we used LED illumination to get single electron
signals with reasonable rate (several kHz) for all further measurements. Figure ~\ref{fig:R11065spectrum} compares pulse height spectra from LED signals with those
obtained from spontaneous emission at room temperature. We can see that both spectra are rather similar and the one corresponding to LED illumination does not
contain any double-electron peak. SER spectra obtained with LED illumination at room temperature and at 87 K are also compared in Fig.~\ref{fig:R11065spectrum}. The spectra
shapes are similar and the noise level is independent of temperature. The gain is higher at low temperature at a fixed anode voltage and this effect takes place for all tested
PMTs (see below). The noise level in all the measurements was determined by radio-frequency pick-ups. The estimate of the first dynode gain, using data on secondary
electron emission from ref. ~\cite{Hamamatsu_Handbook}, gives the value of 7 to 9 at a total PMT gain of 10$^6$ to 10$^7$ correspondingly. In principle, the noise in 
Fig.~\ref{fig:R11065spectrum} can be explained by the signals from thermal single electrons emitted from the first dynode, assuming that its amplification factor is 9. However, during the
measurements the noise level did not change with the PMT gain, as it should be if single electrons from the first dynode are responsible for this.  

\begin{figure}[htb]
\centering
\includegraphics[width=1.0\textwidth,viewport=1 1 750 550,clip]{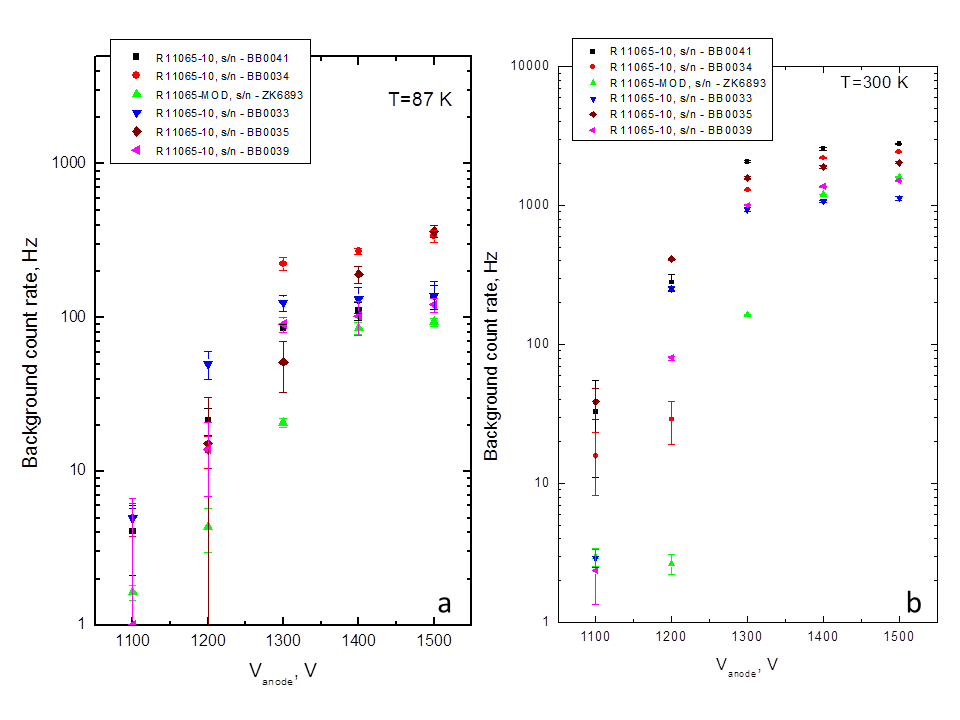}
\caption{Dark count rate as a function of anode voltage for five R11065-10 and one R11065-MOD PMT: a - liquid Ar temperature, b - room temperature.}
\label{fig:noiseR11065}
\end{figure}

Two of the seven tested R11065-10 tubes stopped
functioning after operating for less than an hour at
liquid Ar temperature. The results of dark count rate measurements for the
remaining five R11065-10 tubes and one R11065-MOD tube are presented in Fig.~\ref{fig:noiseR11065}. The discriminator threshold was tuned such that
the counting rate with zero voltage on a PMT did not exceed 1 Hz and then kept constant during all the measurements. 

The gain of the PMT was calculated from the measurements of the pulse height spectra of SER. At first the waveform was measured for each PMT: by integrating it and
expressing the total charge in electrons, the relationship between the full charge and the pulse height was calculated. Then the pulse height spectra were measured and the
most probable pulse height was found for each spectrum by fitting the Gaussian distribution (as in Fig.~\ref{fig:R11065spectrum}). The pulse height distribution of SER in PMT in the
absence of external noise sources is
mainly determined by multiplication statistics of the first dynode that is well described by Poisson distribution ~\cite{statisticsSER}. Poisson distribution with
the average value of 7 to 9 is very close to that of  Gaussian. Moreover, the rest dynodes and electronic noise smear the final distribution following in average the Gaussian law.
Practically
all measured SER spectra of R11065 PMTs had the peak to valley ratio much better than 2 (see Fig.~\ref{fig:R11065spectrum}) and the Gauss function was well fitted 
to the whole distribution but not only to the right slope. 

\begin{figure}[htb]
\centering
\includegraphics[width=0.7\textwidth,viewport=1 300 500 750,clip]{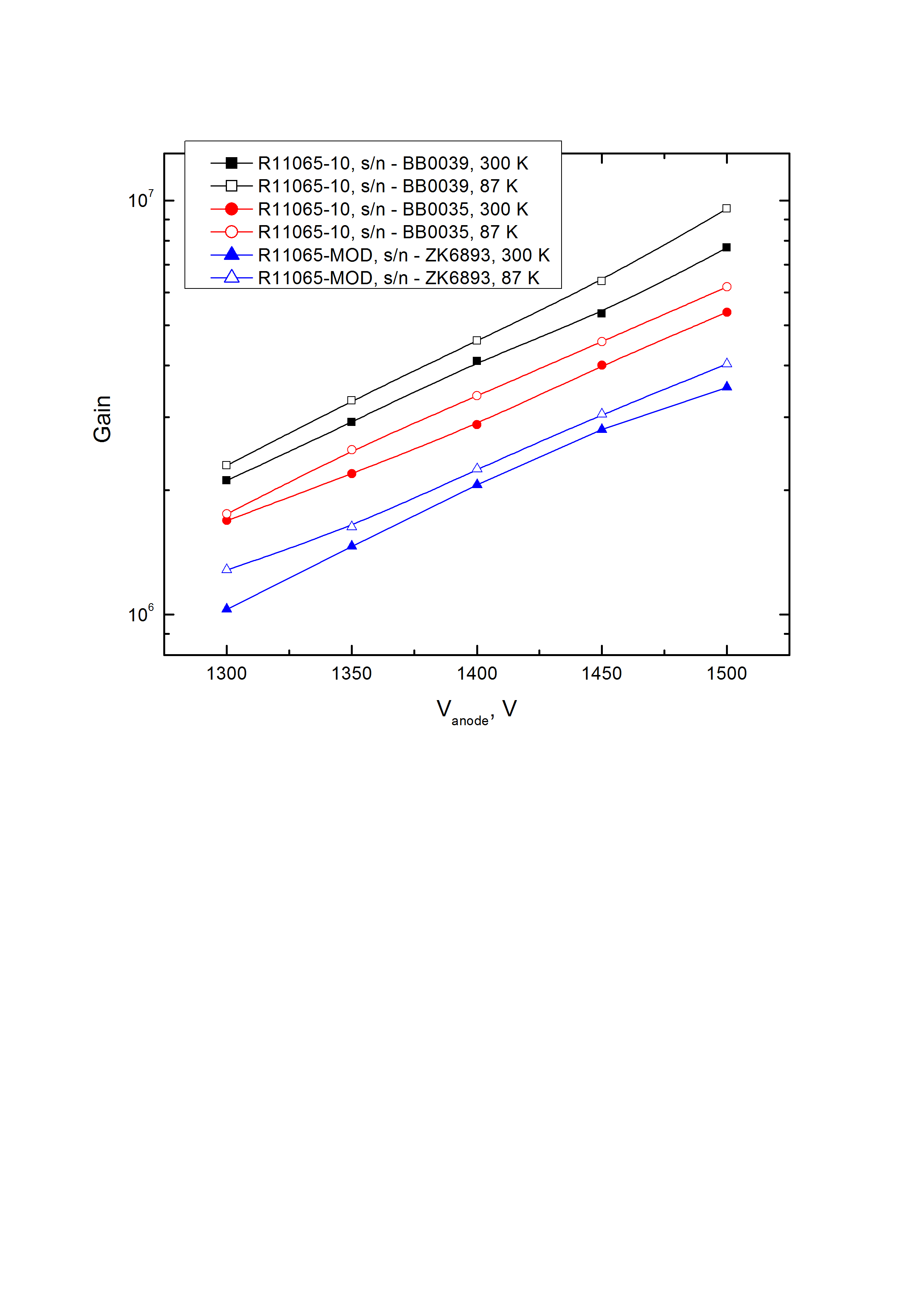}
\caption{Gain as a function of anode voltage for two R11065-10 (one with the highest maximum gain and one with the lowest maximum gain) and one R11065-MOD PMT. 
Results for different PMTs are marked with different colors. Data
corresponding to liquid Ar temperature are marked with open points. }
\label{fig:GainR11065}
\end{figure}

The gain as a function of anode voltage for three R11065 PMTs is shown in Fig.~\ref{fig:GainR11065}. At the highest anode voltage 
of 1500 V the gain at liquid Ar temperature is within the range between 4x10$^6$ (for R11065-MOD) and 10$^7$. 
For R11065-10 the gain at 1500 V is between 5x10$^6$
and 10$^7$. For all the PMTs the gain at low temperature is 10-20\% higher than at room temperature. 

\begin{figure}[htb]
\centering
\includegraphics[width=0.8\textwidth,viewport=1 1 750 550,clip]{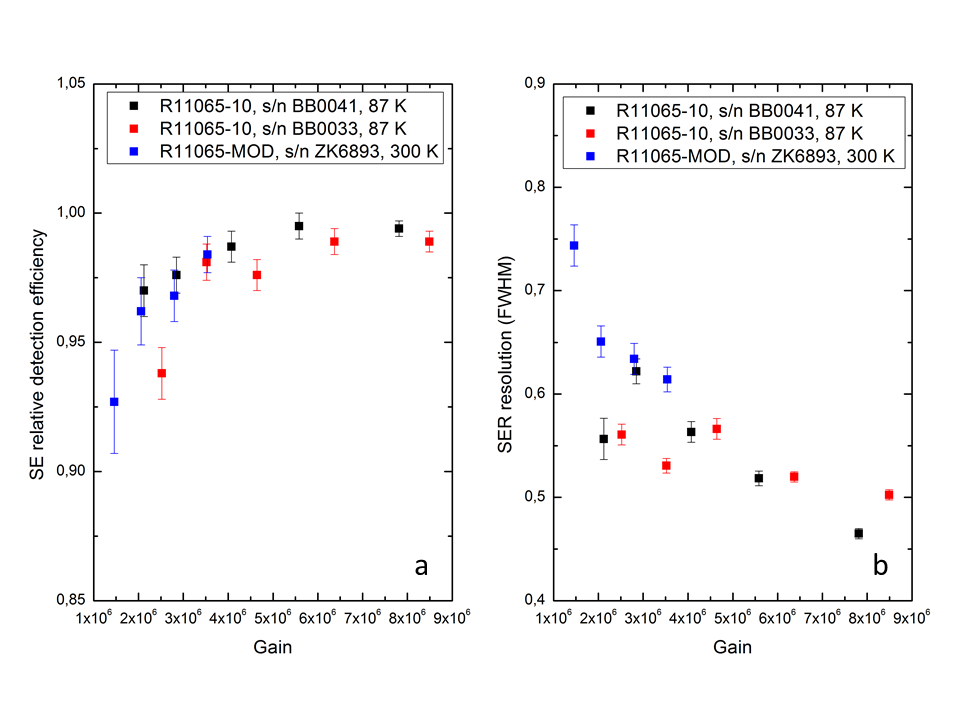}
\caption{Relative detection efficiency of single electrons (a) and pulse height resolution of SER (b) in R11065 as a function of
PMT gain. The data are shown for three PMTs with the lowest, medium and the highest maximum gain.}
\label{fig:R11065resolution}
\end{figure}

The resolution and
the relative single electron detection efficiency as functions of gain are shown, for three PMTs with different maximum gains in Fig.~\ref{fig:R11065resolution}. 
The pulse height resolution is
calculated here as the ratio between FWHM of Gaussian fit and the mean value of the same fit. 
In the figure we can see that the resolution is improved with gain, reaching $\sim$50\% value. As a measure of fraction of single electron signals that can be detected
above a certain threshold, the relative single electron detection efficiency is introduced. This value is calculated as a ratio between an integral of the Gaussian fit above
the threshold and the full integral of the Gaussian fit. The threshold is put at the minimum of the fit to pulse height spectrum below the most probable value as shown in
Fig.~\ref{fig:R11065spectrum}. Due to the rather good pulse
height resolution the relative efficiency comes to plateau at gains above 5x10$^6$ and reaches the value above 99\%.  

\subsection{Results with R6041-506MOD-ASSY PMT.}

\begin{figure}[htb]
\centering
\includegraphics[width=0.7\textwidth,viewport=50 20 800 550,clip]{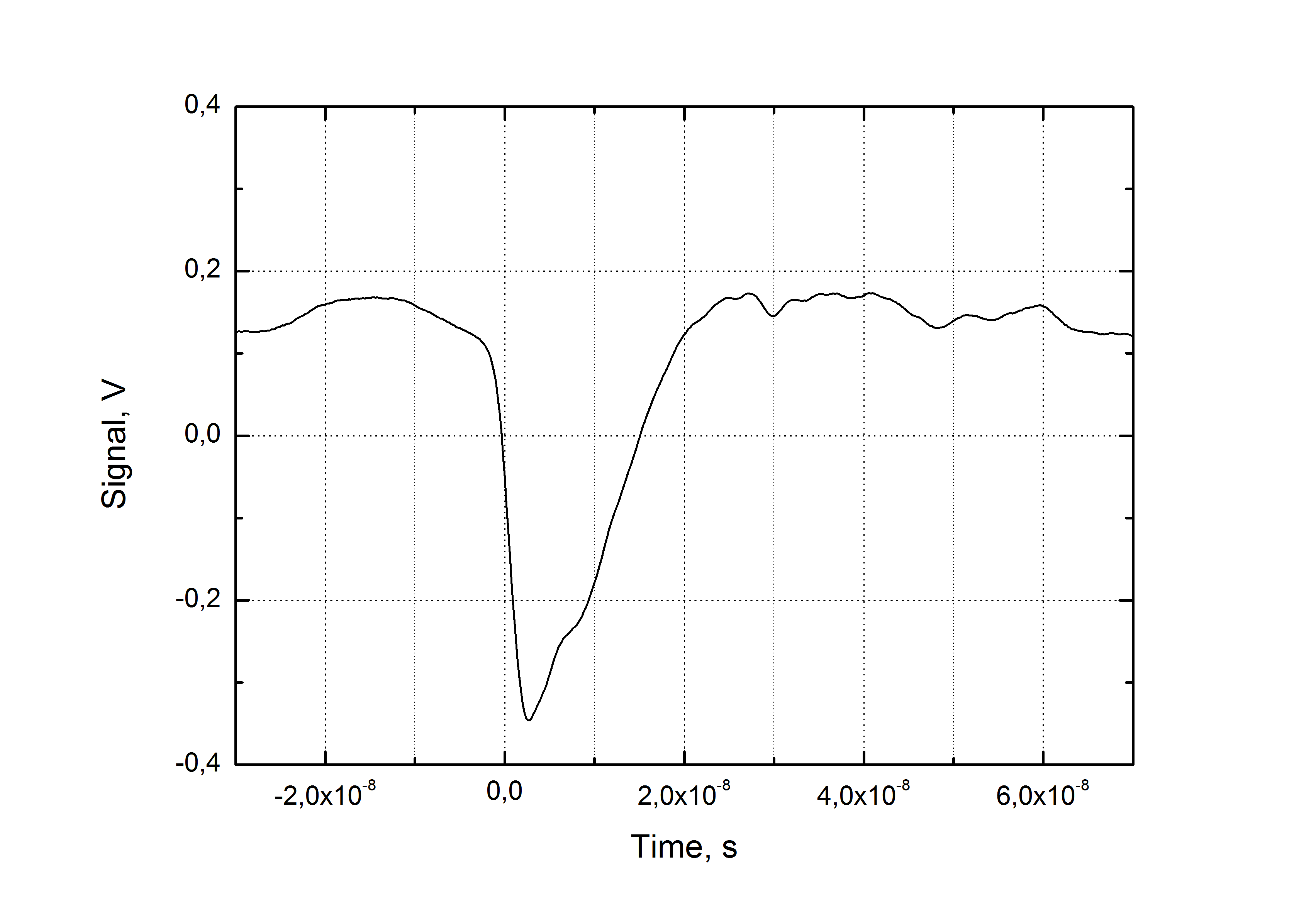}
\caption{Typical signal from R6041-506MOD-ASSY PMT. Anode voltage 900 V, gain $\sim$10$^7$. The signal is amplified by a factor of 100.}
\label{fig:R6041shape}
\end{figure}

\begin{figure}[htb]
\centering
\includegraphics[width=1.0\textwidth,viewport=1 1 750 550,clip]{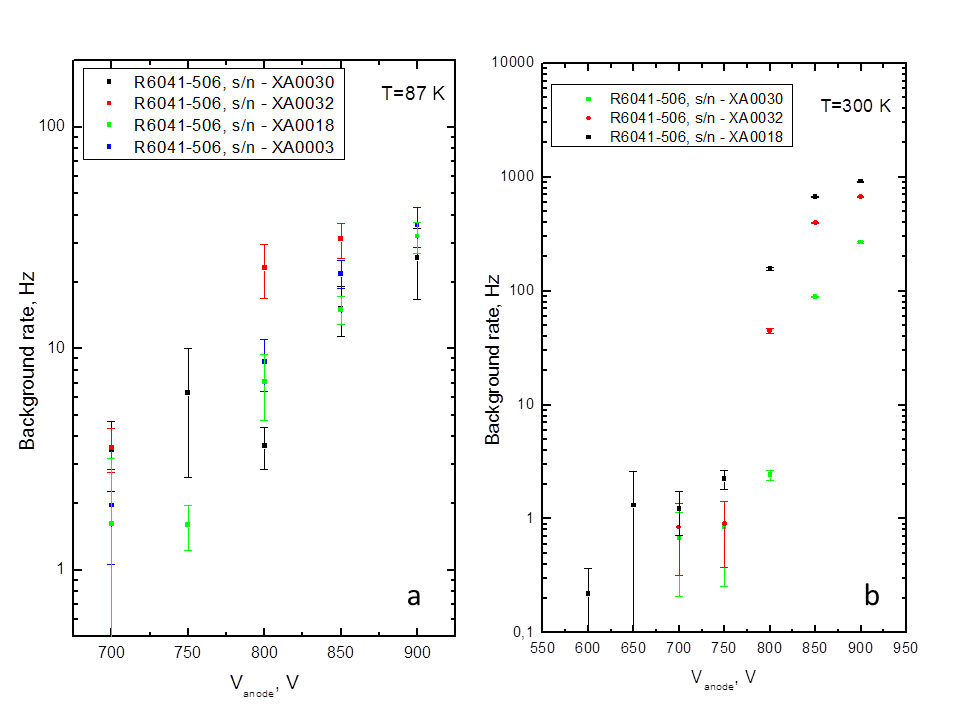}
\caption{Dark count rate as a function of anode voltage for 4 R6041-506MOD-ASSY PMT: a - liquid Ar temperature, b - room temperature.}
\label{fig:R6041noise}
\end{figure}

The typical signal from R6041-506MOD-ASSY is shown in Fig.~\ref{fig:R6041shape}. This PMT is about twice slower than R11065: FWHM of the waveform is $\sim$14 ns
and the width at 0.1 of maximum is about 20 ns. 
As R6041-506MOD-ASSY is slower PMT as compared to R11065, it has approximately twice lower average SE signal at the same gain. 
\begin{figure}[htb]
\centering
\includegraphics[width=0.7\textwidth,viewport=100 1 720 520,clip]{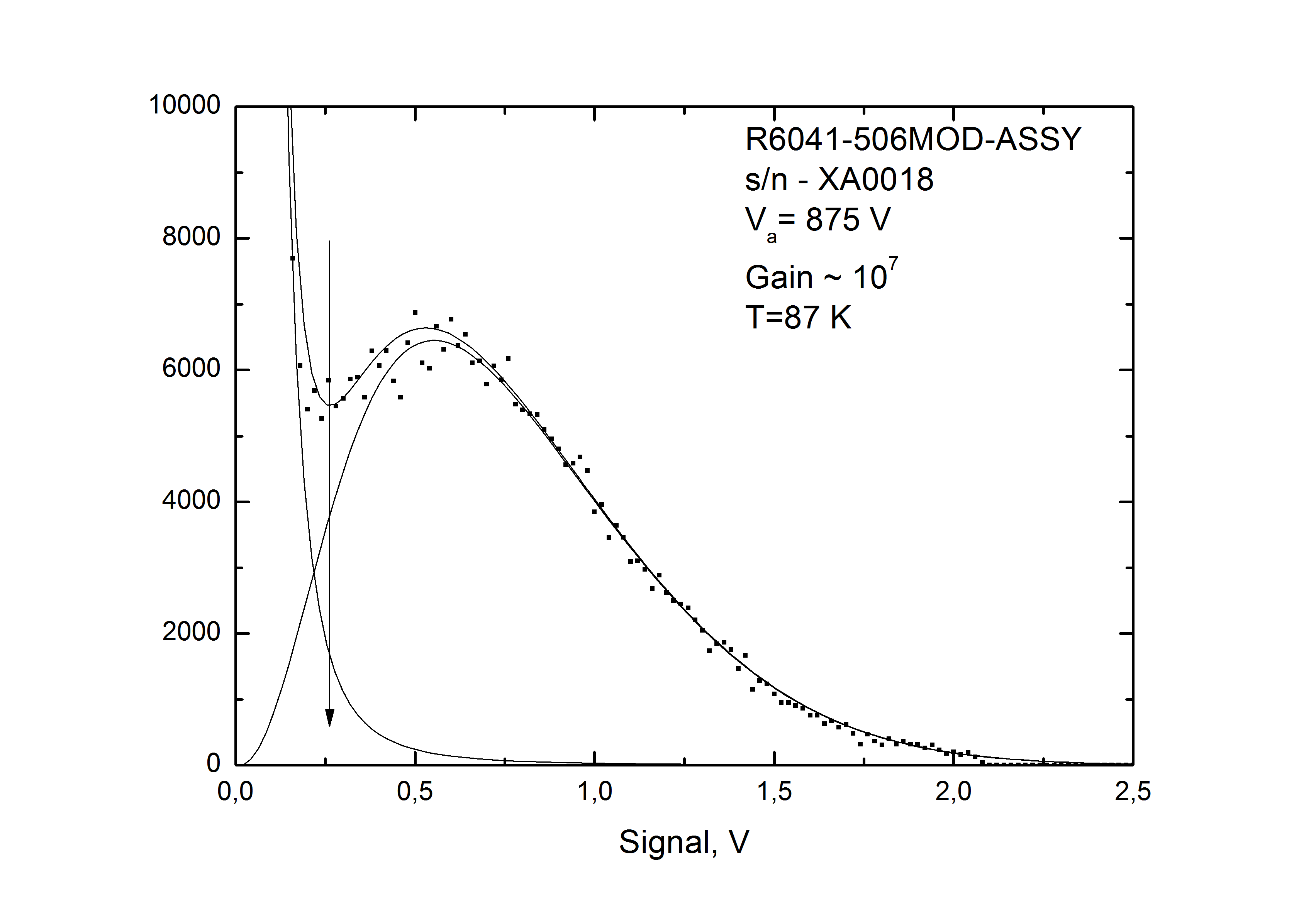}
\caption{Pulse height spectrum of SER of R6041-506MOD-ASSY PMT. Anode voltage 875 V, gain $\sim$10$^7$. Addition amplification of factor 100 is applied. Vertical arrow points
to the threshold for calculation of the relative single electron detection efficiency.}
\label{fig:R6041spectrum}
\end{figure}
Dark count rate in this PMT is significantly lower than in R11065 (Fig.~\ref{fig:R6041noise}). At room temperature the dark count rate does not exceed 1 kHz
at anode voltage of 900 V (the maximum recommended by the manufacturer is 1000 V). 
At liquid Ar temperature the dark count rate 10-30 times lower and does not exceed 50 Hz.

\begin{figure}[htb]
\centering
\includegraphics[width=0.7\textwidth,viewport=50 1 750 550,clip]{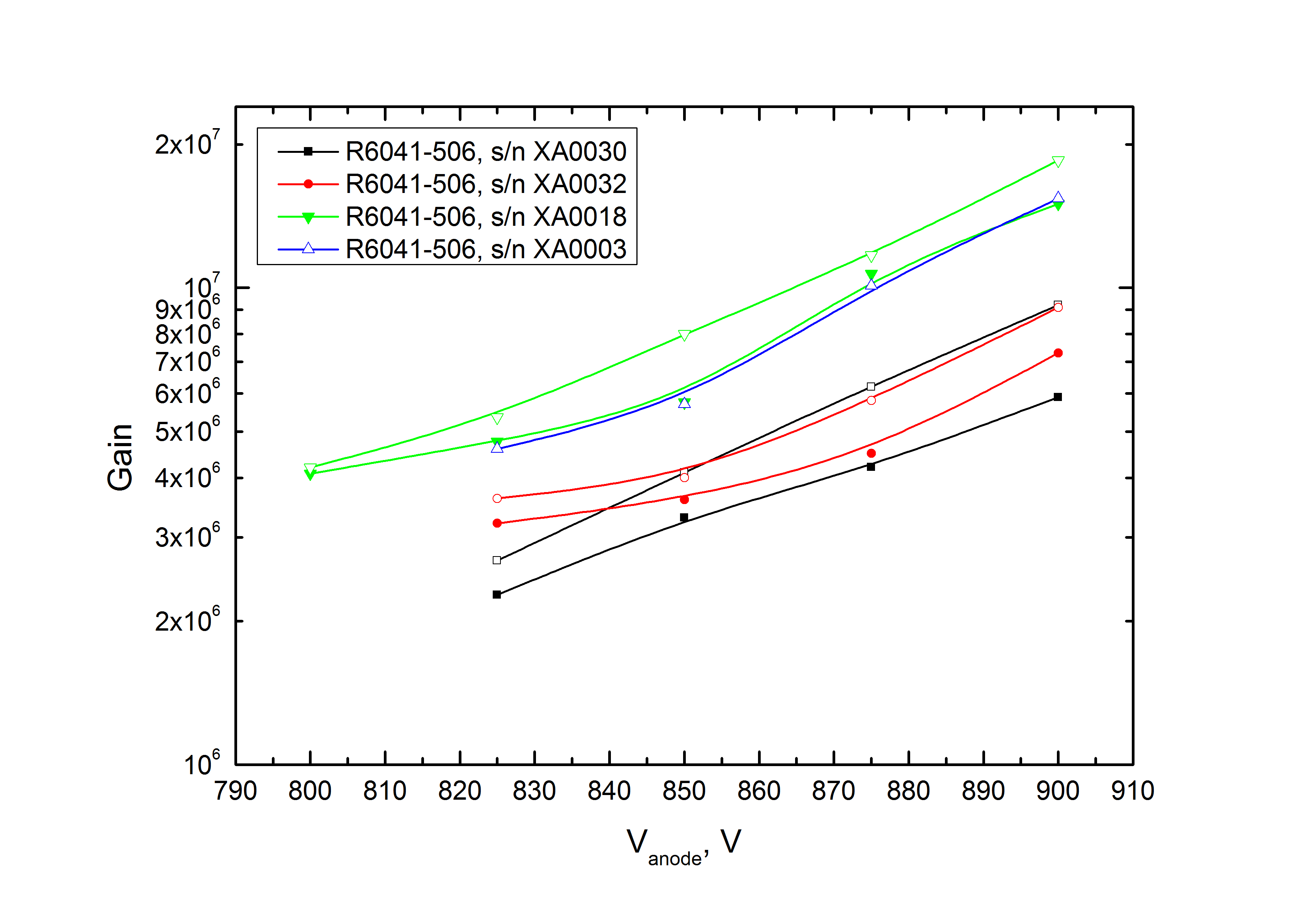}
\caption{Gain as a function of anode voltage for 4 R6041-506MOD-ASSY PMT. Results for different PMT are marked with different colours. Data
corresponding to liquid Ar temperature are marked with open points.}
\label{fig:GainR6041}
\end{figure}

The gain for these PMTs was calculated from the measurements of waveforms and pulse height spectra in a similar way as for R11065 (Fig.~\ref{fig:GainR6041}). For calculation of
the gain we used the average value of the fitted distribution but not the most probable value as for R11065. The estimate of the
first dynode gain, using the data from ref. ~\cite{Hamamatsu_Handbook}, gives the value of 4.5 to 5.5 for total gain of 10$^6$ to 10$^7$. For fitting the SER distributions
we used continuous function that follows Poisson distribution convolved with the Gaussian function: 
\begin{equation}
\centering
y = A\frac{m^{\alpha x} e^{-m}}{\Gamma(\alpha x)}\otimes Gauss(\sigma, \alpha x)
\end{equation}

where $\Gamma$ is the Gamma function, Gauss is the Gaussian function and A, $\alpha$ and $\sigma$ are the free parameters. The noise level was determined by radio frequency
pick ups and was much higher than the level of SER signals from the first dynode. 
 The
gain in the range between 6x10$^6$ and 2x10$^7$ can be obtained for all tested PMTs of R6041-506MOD-ASSY at 900 V. Similarly to R11065 tubes, the gain values for R6041 at
87~K are 10-20\% higher than at room temperature.  

\begin{figure}[hb]
\centering
\includegraphics[width=0.8\textwidth,viewport=1 1 750 550,clip]{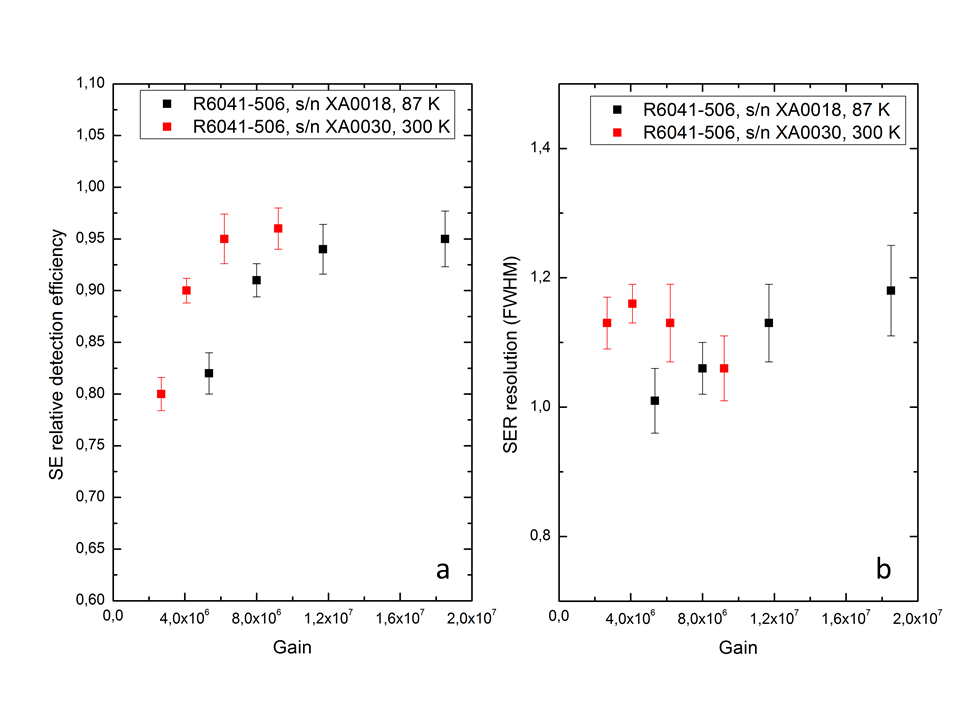}
\caption{Relative detection efficiency of single electron (a) and pulse height resolution of SER (b) in R6041-506MOD-ASSY as a function of
PMT gain for two PMTs with the lowest and the highest maximum gains.}
\label{fig:R6041resolution}
\end{figure}

The R6041-506MOD-ASSY tube  has lower the first dynode gain than in R11065 tube, that results in worse
pulse height resolution of SER as well as lower relative single electron detection efficiency. The resolution was determined as the ratio of 
FWHM and mean value of the fitted distribution. The resolution and relative single electron efficiency for two R6041-506 PMTs with the highest and the lowest 
maximum gains are shown in Fig.~\ref{fig:R6041resolution}. We can see that the resolution is in average close to 1.1. 
The relative single electron detection efficiency is growing with gain and come to plateau at a gain of $\sim$6x10$^6$. The value of the relative SE efficiency on plateau is
around 95\%. 

\section{Conclusions.}

Low background cryogenic PMTs of R11065-10, R11065-MOD and R6041-506MOD-ASSY type, produced by Hamamatsu, were tested when fully immersed in liquid Ar. Devices of R11065 family
demonstrated excellent performance in terms of SER resolution ($\sim$50\% FWHM) and relative single electron detection efficiency (99\%). 
SER spectra and SE relative efficiency were for the first time measured for R6041-506MOD-ASSY PMTs immersed in liquid Ar. 
These PMTs have lower multiplication
at the first dynode than R11065 
and therefore they have worse pulse height resolution ($\sim$110\% FWHM) and relative single electron detection efficiency ($\sim$95\%). 
The relative single electron detection
efficiency depends on particular discriminator threshold that was set to minimize the noise. 
Therefore these values are specific for our particular set-up and noise sources in and
around the laboratory. 
We consider that more than 90\% of single electron detection efficiency is enough for successful operation of these devices in the two-phase CRAD and this value can be obtained
for all types of investigated PMTs even in more noisy conditions.  
All PMTs demonstrated gain between 4x10$^6$
and 2x10$^7$ at the highest anode voltages applied. The dark count rate did not exceed 400 Hz for R11065 and 50 Hz for R6041 PMTs at liquid Ar temperature. In spite of
excellent performance of R11065 devices, we observed breakdown in two R11065-10 tubes and returned them to the manufacturer for exchange. As far as we know the company is
aware of this problem and considering the way to solve it ~\cite{PMTLAr1}.

\section{Acknowledgements}

 This study was supported by Russian Science Foundation (project N 14-50-00080).

\end{document}